\documentclass[twocolumn,preprintnumbers,amsmath,amssymb,prb,aps,superscriptaddress,longbibliography]{revtex4-1}
\usepackage{txfonts}
\usepackage{pifont}
\usepackage{amssymb}
\usepackage{dcolumn}
\usepackage{amsmath}
\usepackage[dvips]{epsfig}
\usepackage{array}
\usepackage{color}
\usepackage{ulem}
\usepackage{caption}
\usepackage{graphicx}
\usepackage{float}
\usepackage{subcaption}
\usepackage{titlesec}
\usepackage[colorlinks=true,linkcolor=blue,citecolor=blue,urlcolor=blue,]{hyperref}
\newcolumntype{C}[1]{>{\centering}p{#1}}
\makeatletter
\def\btt#1{\texttt{\@backslashchar#1}}%
\DeclareRobustCommand\bblash{\btt{\@backslashchar}}%
\makeatother

\begin{document}

\title{Ferroelectricity originating from polar moiety flipping}

\author{Zengqian Wang}
\affiliation{College of Physics and Engineering, Qufu Normal University, Qufu, Shandong 273165, China}
\author{Yuanfang Yue}
\affiliation{College of Physics and Engineering, Qufu Normal University, Qufu, Shandong 273165, China}
\author{Z. Y. Xie}
\affiliation{School of Physics, Renmin University of China, Beijing 100872, China}
\affiliation{Key Laboratory of Quantum State Construction and Manipulation (Ministry of Education), Renmin University of China, Beijing 100872, China}
\author{Fengjie Ma}
\email{fengjie.ma@bnu.edu.cn}
\affiliation{The Center for Advanced Quantum Studies and School of Physics and Astronomy, Beijing Normal University, Beijing 100875, China}
\affiliation{Key Laboratory of Multiscale Spin Physics (Ministry of Education), Beijing Normal University, Beijing 100875, China}
\author{Xun-Wang Yan}
\email{yanxunwang@163.com}
\affiliation{College of Physics and Engineering, Qufu Normal University, Qufu, Shandong 273165, China}

\date{\today}

\begin{abstract}

A wide variety of applications has inspired great interest in designing new materials and investigating fundamental physics in the ferroelectric field. In the concept of ferroelectricity, the spontaneous polarization is traditionally considered a core property, but we find that the `spontaneous' seems unnecessary. Based on this understanding, we suggest a new type of ferroelectric materials, in which the polar atomic moiety is incorporated to realize the electric polarization and the polar atomic moiety flipping corresponds to the reversal of electric polarization. Lead hydroxyapatite Pb$_{10}$(PO$_4$)$_6$(OH)$_2$ containing polar OH moieties is taken as an example to illustrate our idea, in which the OH moiety flipping result in the reversal of electric polarization.
The mechanism of ferroelectricity reported herein distinctly differs from those of the displacive type, disorder-order type, and interlayer sliding type ferroelectric materials in the previous studies.

\end{abstract}

%\pacs{74.70.Kn, 74.20.Pq, 61.66.Hq, 61.48.-c}

\maketitle
%% introduction of 2D materials, like 2016 PRB
%% manipulation the electronic structure, tailor elocate
%% magnetic 2D is rare
%% Here, magneism change from Cr - Mn Fe Ni

%\section{Introduction}

\textit{Introduction}.---Ferroelectricity refers to the phenomenon of spontaneous electric polarization existing in certain materials, which can be reversed by the applied electric field and has a hysteresis effect with respect to the applied electric field. The spontaneous polarization as well as the hysteresis effect can be used as a memory function to make nonvolatile memory devices\cite{Dawber2005,Scott1989,Whatmore2017}. Capacitors made of ferroelectric materials have the characteristic of adjustable capacitance, which are the main component of some sensors, e.g. temperature sensors, sonar, and vibration sensors\cite{Scott2007}. Ferroelectric tunneling function is another application with a giant electroresistance switching effect\cite{Zhuravlev2005}. Moreover, when ferroelectric ordering and ferromagnetic ordering coexist in material or heterostructure, the multiferroics have potential for applications as actuators, switches, magnetic field sensors, and new types of electronic memory devices\cite{Ramesh2007}.
A wide variety of applications has inspired great interest in designing new materials and investigating fundamental physics in this field\cite{PhysRevLett.133.246703,PhysRevLett.132.256801}.

 A typical ferroelectric of the displacive type is BaTiO$_3$, in which the Ti ion is displaced from equilibrium and the oxygen octahedral cage is distorted. The asymmetrical shift of Ti ion in the equilibrium ion positions leads to a permanent dipole moment and hence to a ferroelectric phase\cite{Devonshire1949}.
 KH$_2$PO$_4$ is known as a prototype of order-disorder ferroelectric materials, in which different possible arrangements of the hydrogens result effectively in different orientations of the (H$_2$PO$_4$)$^-$ dipoles.\cite{Slater1987}
Recently, Wu \textit{et al.} proposed interlayer sliding ferroelectricity\cite{Li2017a}, which is induced by specific stacking modes of two van der Waals (vdW) layers and can be reversed by interlayer sliding.
According to the concept of ferroelectricity, the electric polarization is spontaneous in the compounds below the Curie temperature. But for the recently discovered sliding ferroelectricity, the word `spontaneous' seems superfluous, because the electric polarization is permanent when the two layers are arranged in the `AB' pattern.
Therefore, two factors are the core of ferroelectricity: one is the existence of electric polarized state, and the other is the reversal between two opposite polarized states.
Based on the above analysis, we have clarified the concept of ferroelectricity and the formation mechanism of displacive, disorder-order, and sliding ferroelectric materials.
Now, the question we are thinking about is whether there exist new ways to realize the ferroelectricity.

Let's focus on polar molecules. Most of molecules are polar, such as H$_2$O and NH$_3$. In them, the center of positive and negative charges in the molecule does not coincide, which leads to an electric dipole in the molecule. So, the polar molecule is naturally in a polarized state. If we manually flip the molecule by 180 degrees and make the dipole moment along the opposite direction, the reversal of the polarized state is realized. For such a polar molecule, it possesses the two key factors of ferroelectricity, polarized state and the reversal. But this is just a microscopic single molecule, not a macroscopic bulk material or a macroscopic two-dimensional material.
Considering the dipole moment of some polar molecules or moieties, we can incorporate or intercalate the polar molecules or moieties into the bulk compounds with large interstitial space to design new ferroelectric materials. In this work, the lead hydroxyapatite Pb$_{10}$(PO$_4$)$_6$(OH)$_2$ with a polar OH moiety incorporated is taken as an example to reveal our design idea.

%\section{Computational details}
\textit{Computational details}.---The plane wave pseudopotential method enclosed in the Vienna \textit{ab initio} simulation package (VASP) \cite{PhysRevB.47.558, PhysRevB.54.11169}, along with the pseudopotential based on the projector augmented-wave method (PAW) \cite{PhysRevB.50.17953} and the Perdew-Burke-Ernzerhof (PBE) functional \cite{PhysRevLett.77.3865} were adopted in the calculations.
The energy cutoff for the plane wave basis was 600 eV and the convergence thresholds of total energy and force was 10$^{-5}$ eV and 0.01 eV/\AA ~.
A mesh of $12\times 12\times 12$ k-points were sampled for the Brillouin zone integration in the self-consistent electronic structure calculations.
The climbing image nudged elastic band (CINEB) method\cite{Henkelman2000a} is used to search the minimum energy path and estimate the ferroelectric switching energy barrier, and the Berry phase approach\cite{King-Smith1993,Resta1994} is applied to evaluate the electric polarization.

%\section{Results and discussion}
\textit{Results and discussion}.---In July 2023, the claimed superconductivity in Cu-doped lead apatite Pb$_{10-x}$Cu$_x$(PO$_4$)$_6$O with $0.9 \leq x \leq 1$  (dubbed as LK-99) \cite{Lee2023,Lee2023a} with the critical temperature at 400 K was reported by Lee $et ~al.$ Subsequently, the research on the copper-substituted lead apatite was booming, including the synthesis of the samples, measurement of various physical properties, theoretical simulation of the electronic structure, and analysis of possible superconductivity\cite{Liu2024a,Jiang2023a,He2023,Cabezas-Escares2024}. Up to date, no research group can replicate the zero resistance and the magnetic levitation from the diamagnetism simultaneously. Because of the considerable debate and controversy about superconductivity\cite{Wang2024}, lead apatite compounds have aroused great concern within the scientific community. We focus on the ferroelectricity of lead hydroxyapatite, which has seldom been investigated in the apatite family in the past.

\begin{figure}[htp]
\begin{center}
\includegraphics[width=8.0cm]{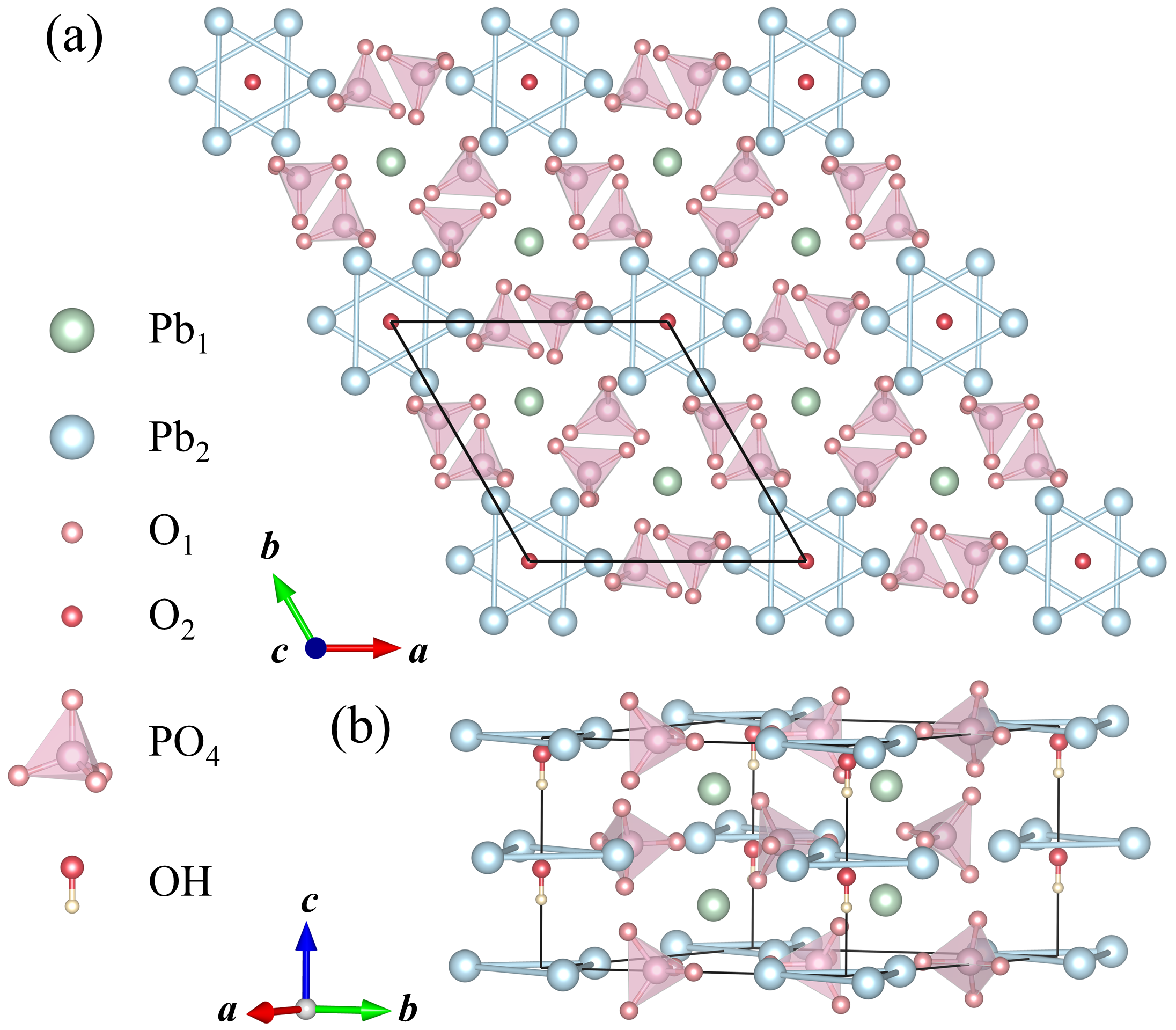}
\caption{Atomic structure of Pb$_{10}$(PO$_4$)$_6$(OH)$_2$, (a) 2 $\times$ 2 $\times$ 1 supercell in the top view; (b) single unit cell in the side view. The unit cell is marked with a black solid line box. The green, blue, pink, red, lavender and yellow spheres represent Pb$_1$, Pb$_2$, O$_1$, O$_2$, P, and H atoms, respectively.
 } \label{structmodel}
\end{center}
\end{figure}

%\subsection{Atomic structure}

Lead hydroxyapatite is a mineral with a hexagonal structure and its chemical formula is Pb$_{10}$(PO$_4$)$_6$(OH)$_2$. It belongs to the space group $P6_3$ (No.173), and the lattice parameters are  $a$ = 9.866 \AA~ and $c$ = 7.426 \AA~\cite{Brtickner1995,White2003}.
 Fig.~\ref{structmodel} shows the atomic structure of Pb$_{10}$(PO$_4$)$_6$(OH)$_2$.
In the structure of lead hydroxyapatite, Pb and O atoms can be distinguished into Pb$_1$, Pb$_2$, O$_1$, and O$_2$ sites depending on their site symmetry. Pb$_2$ atoms sit in the $6h$ Wyckoff site and every three Pb$_2$ atoms form a triangle located at the edge of unit cell. Four O$_1$ atoms and one P atom form a tetrahedral PO$_4$ unit, and in the $ab$ plane the PO$_4$ tetrahedrons are connected by Pb$_2$ atoms while along the $c$ axis these PO$_4$ tetrahedrons are connected by another kind of Pb atom at the $4f$ Wyckoff site, marked as Pb$_1$. The PO$_4$ tetrahedrons and Pb atoms make up the basic framework of the lead hydroxyapatite.
O$_2$ atoms are aligned linearly along the \textit{c}-axis and are located at the centers of Pb$_2$ triangles, or in other words, OH moiety is in a one-dimensional channel enclosed by Pb$_2$ atoms.

\begin{figure}[htp]
\begin{center}
\includegraphics[width=8.0cm]{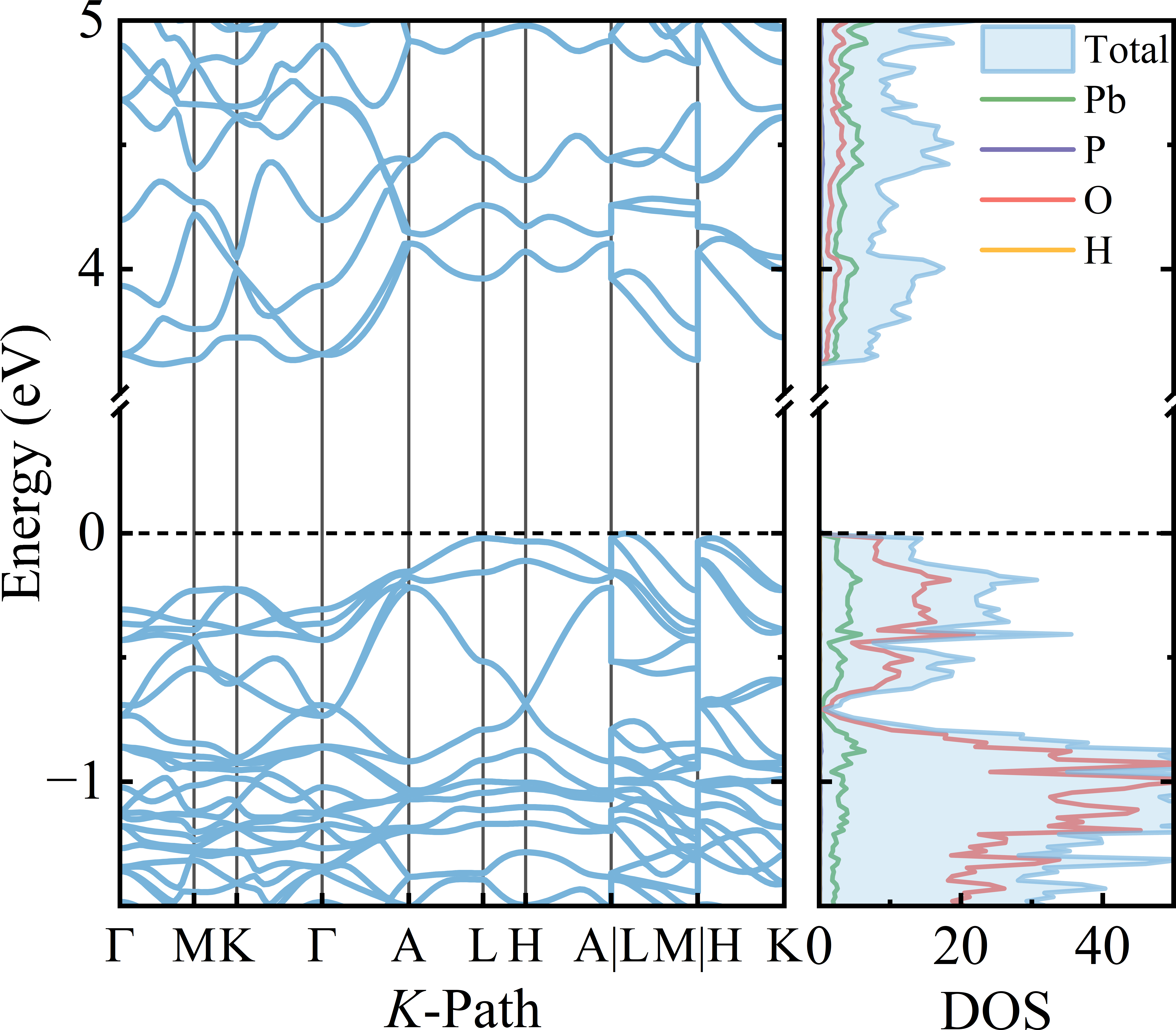}
\caption{The band structure and density of states of Pb$_{10}$(PO$_4$)$_6$(OH)$_2$ in the ferroelectric phase.
} \label{fig2-band}
\end{center}
\end{figure}

%\subsection{Electronic structure}
The band structure and density of states of Pb$_{10}$(PO$_4$)$_6$(OH)$_2$ are shown in Fig.~\ref{fig2-band}, which indicate that Pb$_{10}$(PO$_4$)$_6$(OH)$_2$ is a non-magnetic insulator.  In Fig.~\ref{fig2-band}, the green, purple, red and orange lines represent the density of states projected onto the Pb, P, O, and H atoms, respectively. The electron states near the Fermi level are mainly contributed by Pb and O. The band gap of Pb$_{10}$(PO$_4$)$_6$(OH)$_2$ is 3.62 eV, which is a relatively large band gap. The energy band gap is a key parameter for the application of ferroelectric materials, because the large band gap can reduce the generation of charge carriers and minimize the leakage current, thereby improving the device efficiency and stability in the high electric field. The well-known ferroelectric materials for practical applications are BaTiO$_3$, SrTiO$_3$, and PbTiO$_3$, whose electronic structures have been studied by the leading scientists such as Cohen and Vanderbilt\cite{Cohen1992,King-Smith1994}. A systematic theoretical study on the band gaps of BaTiO$_3$, SrTiO$_3$, and PbTiO$_3$ was reported by Piskunov $et~ al.$, and they are 1.86 eV, 1.99 eV, and 2.0 eV from the PBE functional calculations \cite{Piskunov2004}. The band gaps of BaTiO$_3$, SrTiO$_3$, and PbTiO$_3$ measured in experiments are 3.2 eV, 3.25 eV, 3.45 eV, respectively\cite{Wemple1970,VanBenthem2001,Peng1991}. In the previous literatures, we do not find any experimental reports on the band gap of Pb$_{10}$(PO$_4$)$_6$(OH)$_2$. The value of 3.62 eV from our calculations indicates that lead hydroxyapatite is an insulator with a larger energy gap than BaTiO$_3$, SrTiO$_3$, and PbTiO$_3$.

\begin{figure}[htp]
\begin{center}
\includegraphics[width=7.0cm]{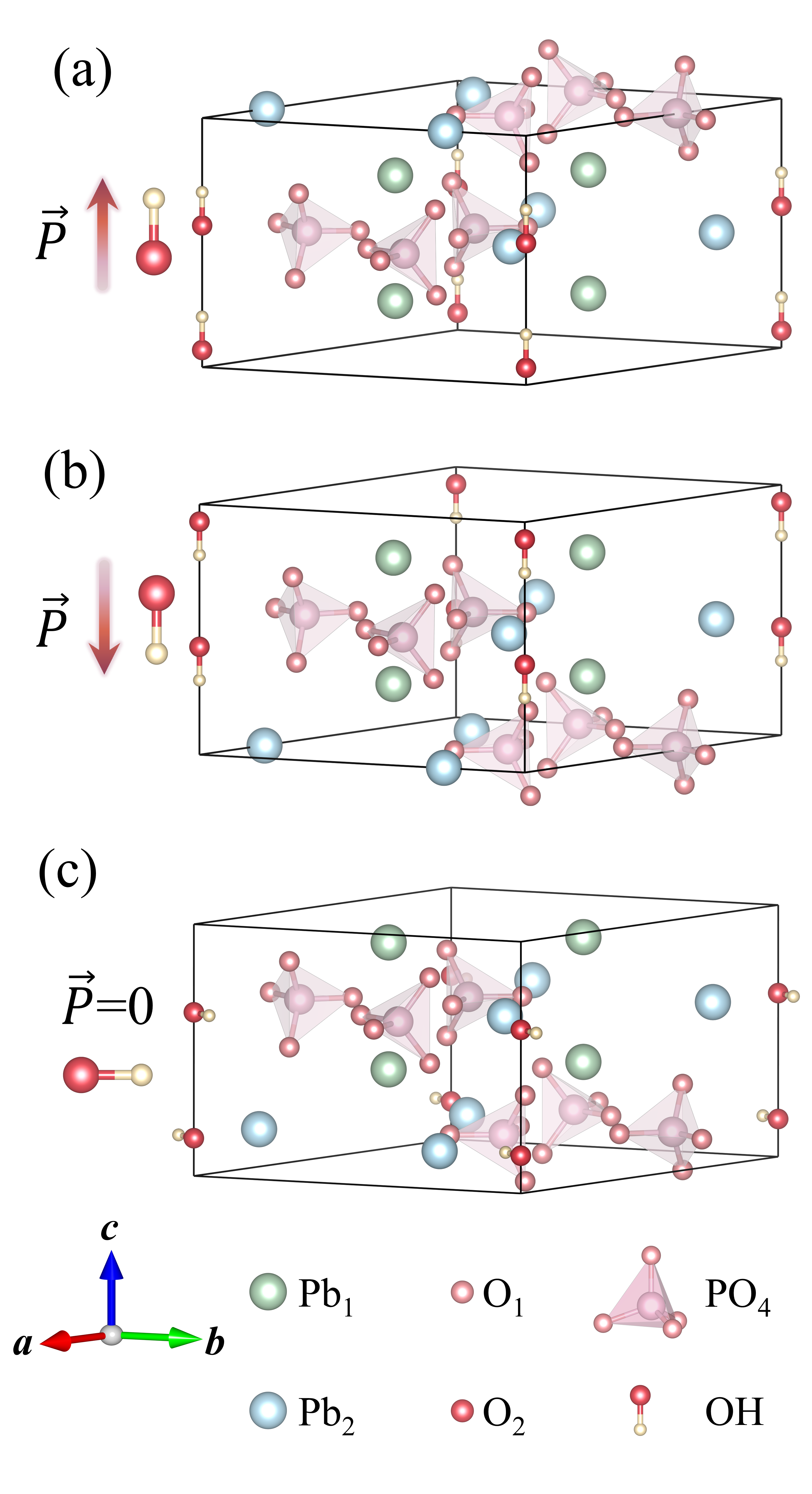}
\caption{Ferroelectricity in Pb$_{10}$(PO$_4$)$_6$(OH)$_2$. (a) up-polarized state, (b) down-polarized state, (c) paraelectric state with the OH moieties parallel to the $ab$ plane.
 } \label{fig3-twop}
\end{center}
\end{figure}

%\subsection{Ferroelectric properties}
%\subsubsection{Electric polarization}

\begin{figure}[htp]
\begin{center}
\includegraphics[width=7cm]{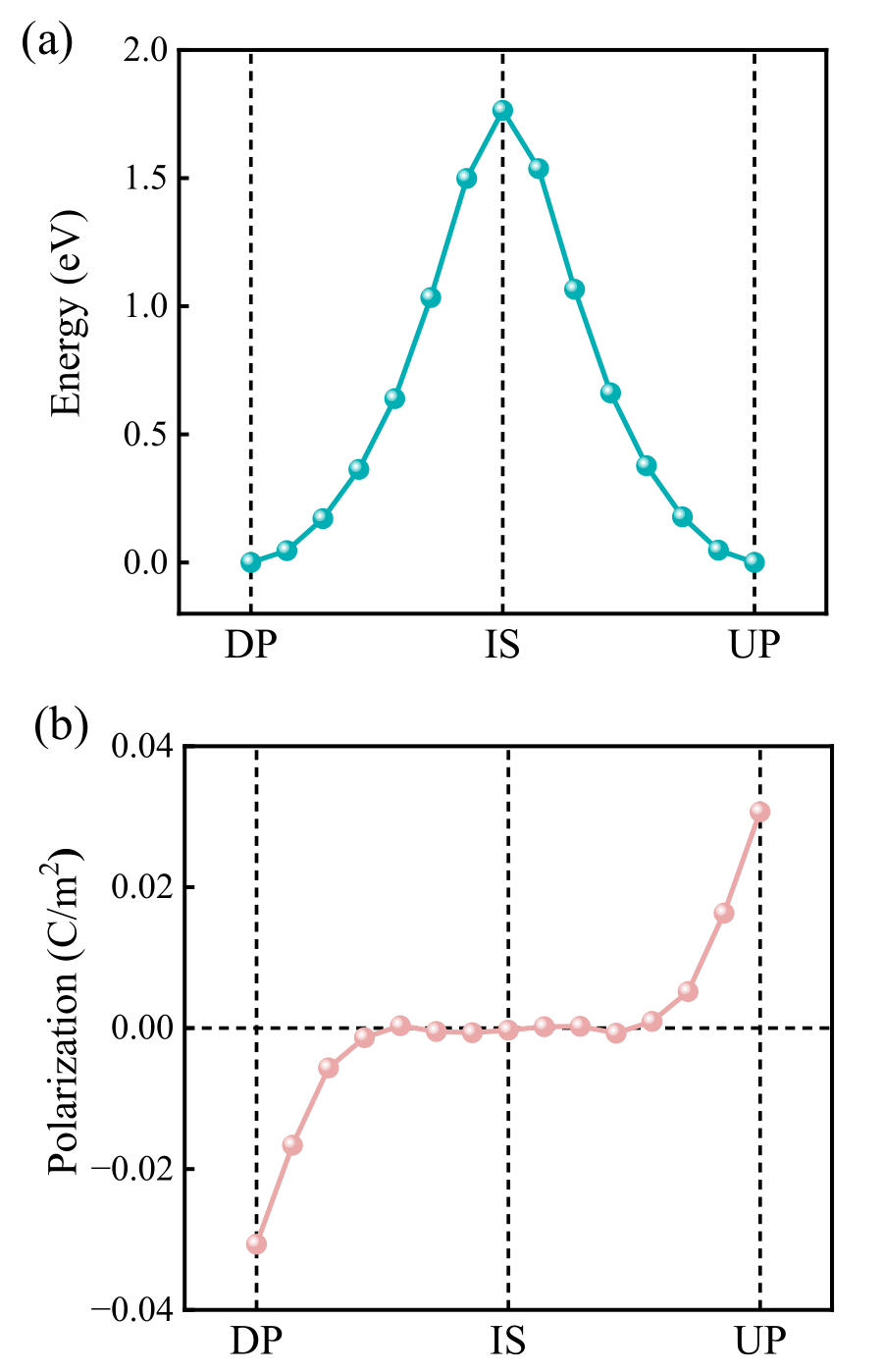}
\caption{(a) Energy barrier and (b) electric polarization varying with the ferroelectric polarization reversal pathway. DP, UP, and IS represent the down-polarized phase, up-polarized phase, and paraelectric phase, respectively.
 } \label{fig4-d-E}
\end{center}
\end{figure}

It is shown in Fig.~\ref{fig3-twop}(a) that the up-polarized phase of Pb$_{10}$(PO$_4$)$_6$(OH)$_2$ with the O$\rightarrow$H pointing to the positive $c$ axis. The value of electric polarization is 0.036 C/m$^2$ and the direction is along the positive $c$ axis. For the down-polarized phase, the direction of O$\rightarrow$H is reversed to point to the negative direction of the $c$ axis, causing the reversal of electric polarization, shown in Fig.~\ref{fig3-twop}(b).
The symmetry of Pb$_{10}$(PO$_4$)$_6$(OH)$_2$ in the two polarized phases belongs to the space group of P6$_3$.
One unit cell of Pb$_{10}$(PO$_4$)$_6$(OH)$_2$ has two OH moieties. When the two OH moieties are rotated by 90 degrees and antiparallelly aligned in the $ab$ plane, the compound changes to the paraelectric phase, as shown in Fig.~\ref{fig3-twop}(c). The space group of the paraelectric phase is P$2_1$/m and it has a spatial inversion symmetry.
The energy difference per formula cell between the ferroelectric phase and the paraelectric phase of Pb$_{10}$(PO$_4$)$_6$(OH)$_2$ are displayed in Fig.~\ref{fig4-d-E}(a). The energy barrier is 1.69 eV per formula cell with the phase transition  from the down-polarized phase to the paraelectric phase, and subsequently to the up-polarized phase. A formula cell contains 44 atoms, and the energy barrier per atom is 38.4 meV.
Fig.~\ref{fig4-d-E}(b) presents the evolution of electric polarization with the structure changing from the down-polarized ferroelectric phase to the up-polarized ferroelectric phase.

%\subsubsection{to confirm electric polarization from OH, compared with the Pb$_{10}$(PO$_4$)$_6$(OH)$_2$}
%\subsubsection{Reversal of electric polarization}
%\subsubsection{Mechanism of ferroelectricity}
As for the mechanism of ferroelectricity in Pb$_{10}$(PO$_4$)$_6$(OH)$_2$, the polar OH moiety is the key component for the origin of electric polarization.
To reveal the critical role of the polar OH moiety, we replace the OH moiety with Cl atom in Pb$_{10}$(PO$_4$)$_6$(OH)$_2$. Although Pb$_{10}$(PO$_4$)$_6$Cl$_2$ has the similar structure to Pb$_{10}$(PO$_4$)$_6$(OH)$_2$, the ferroelectricity do not appear in Pb$_{10}$(PO$_4$)$_6$Cl$_2$.
The fact clearly shows that the ferroelectricity is closely associated to the introduction of the polar OH moiety.
By turning the polar OH moiety, the direction of electric polarization is reversed. Consequently, the inclusion of polar moiety and the flipping are the keys to the origin of ferroelectricity, which is a distinct mechanism of ferroelectricity from those mentioned in the previous studies.

%\section{Discussion and Conclusion}
\textit{Discussion and Conclusion}.---In conclusion, we take Pb$_{10}$(PO$_4$)$_6$(OH)$_2$ as an example to show a new class of ferroelectric compounds, which has not been reported in the previous studies. The compounds contain the polar OH moiety, such as hydroxyapatite and similar compounds with a general formula of apatites M(XO$_4$)(OH), where position M can be occupied by Ca and Pb (also by Zn, Cd, Fe), while the anion XO$_4$ represents PO$_4$, VO$_4$, SO$_4$, and AsO$_4$\cite{1994111}.  In these compounds, the ferroelectric properties result from the polar moiety and the rotation of the moiety, which is a completely new physical picture of the ferroelectric mechanism. On the other hand, we find the `spontaneous' in the concept of ferroelectricity is not necessary because the electric polarization is an inherent characteristic in these compounds.
Furthermore, the polar moiety can also be NH, COOH, CO-NH$_2$, NH$_3$, and H$_2$O, which are included in the parent compounds other than apatites to display the ferroelectricity.
Therefore, our results not only propose a new class of ferroelectric materials with distinct physical mechanism from the known ferroelectric materials but also extend the concept of ferroelectricity.

%%\begin{acknowledgments}
\textit{Acknowledgments}.---This work was supported by the National Natural Science Foundation of China under Grants Nos. 12274255, 12074040, 12274458, 11974207. F. Ma was also supported by the BNU Tang Scholar.
%\end{acknowledgments}

%\bibliography{ref-CrN4C2,Ref}
\bibliography{Ref}

\end{document}